\documentstyle[12pt]{article}
\begin{document}

\begin{center}
{\bf Coherent  tunneling and quantum coherence oscillations \\
at the atomic level\footnote{ Report FT-416-June 1996, 
Institute of Atomic Physics, Bucharest} \\[1cm]}  
{ Marius Grigorescu \\[3cm]  }
\end{center}
\noindent
$\underline{~~~~~~~~~~~~~~~~~~~~~~~~~~~~~~~~~~~~~~~~~~~~~~~~~~~~~~~~
~~~~~~~~~~~~~~~~~~~~~~~~~~~~~~~~~~~~~~~~~~~}$ \\[.3cm]
The evolution of the quantum wave packet describing an atom trapped in the surface-tip junction of the scanning tunneling microscope is investigated by using 
the time-dependent Schr\"odinger equation, and a quasi-classical
Hamiltonian approach. The estimates concern a Xe atom in a 
biased double-well junction potential. The exact treatment 
shows that quantum coherence oscillations of the metastable ground 
state may occur at particular  resonant values of the bias voltage. The 
effect of decoherence by partial localization is studied within the 
quasi-classical frame.  \\
$\underline{~~~~~~~~~~~~~~~~~~~~~~~~~~~~~~~~~~~~~~~~~~~~~~~~~~~~~~~~
~~~~~~~~~~~~~~~~~~~~~~~~~~~~~~~~~~~~~~~~~~~}$ 
\newpage

\section{ Introduction}  
The phenomenology of the
wave function collapse and decoherence is a subject of wide interest,
ranging from  the  conceptual framework of quantum mechanics  
\cite{gm, zurek} to the physics of the quantum  logic circuits. 
\\ \indent
Dynamical loss of coherence produced by possible non-linear terms in the
time-dependent Schr\"odinger equation (TDSE) could be observed 
 by measurements on driven  hyperfine transitions in $^9$Be$^+$ ions 
\cite{wein}.  Such non-linear terms can appear by the coupling to   
the environment \cite{gc}, but they could also be intrinsically 
built into the quantum mechanics, as it was  proposed 
by Ghirardi, Rimini and Weber to explain the paradoxes of the measurement theory  \cite{ghir}. The  mean-field of molecules or atomic 
nuclei provided by the Hartree-Fock approximation 
breaks spontaneously the translation symmetry, and is localized
in the inertial frame defined by the center of mass (CM). 
The CM motion can be quantized, removing the spurious effects of
the symmetry breaking, but when the size  of the objects increases 
towards the macroscopic level, (the object becomes "environment"
for its constituents), physical localization may occur   
\cite{vit, tes}. \\ \indent
The interplay between quantum coherence and wave function collapse 
appears particularly striking in the computation models of the  
quantum computers \cite{shor}. Therefore, a most interesting issue is
to  study the coherence properties of the physical systems at atomic level. 
Recent experiments have proved the existence of a coherent component for
the electron tunneling in quantum dots \cite{qd}, while
 quantum localization was predicted for atoms moving in the
periodic potential created by a phase-modulated standing light
field \cite{ql}.
\\ \indent
The purpuse of this work is to investigate the feasibility of
experiments on localization and  decoherence for the CM wave function
of the individual atoms, using the scanning tunneling microscope (STM).
Since the first experiments on reversible atom 
transfer   \cite{ei}, STM may be considered as the ideal instrument for  
manipulating atoms or molecules. 
The bistable operation mode of STM can be understood  assuming  
that the diffusion barrier on the surface is high enough to prevent
the escape of the particle from the junction region, and that the motion takes
place along the outer normal to the surface plane  in an 
asymmetric, one-dimensional, double-well  potential (DWP) \cite{walk}. 
In a DWP, an observable effect which is very  sensitive to decoherence is 
the tunneling  of a wave-packet $\psi_0$ with  the energy
below barrier \cite{legg}. If $\psi_0$ is a Gaussian localized initially in the metastable well, and the quantum coherence is preserved in time, then complete tunneling to the stable well appears only in  very special  resonance conditions, by  quantum coherence oscillations (QCO) \cite{nieto}. 
\\ \indent
 The coupling to the environment affects the superposition principle and the quantum coherence property, and in the case of a dissipative force linear in the CM velocity, the QCO between the two wells may either be damped \cite{bend}, or completely suppressed \cite{bm}. However, decoherence could be produced also by dissipative terms which  only suppress the spreading of the wave packet (e.g. the  "squeezing"  dissipation \cite{gc}). In the asymmetric DWP, such terms may change the tunneling mechanism from QCO between wave functions with different shapes, to coherent tunneling (i.e. after barrier crossing the wave function retains its initial shape \cite{nieto}). \\ \indent
In Sect. 2 it is shown that for a Xe atom placed in the STM surface-tip junction,
reversible tunneling by QCO resonances could be observed. The irreversible switching mode is discussed in Sect. 3, within a quasi-classical model. Conclusions are summarized in Sect. 4.   
\section{The atomic quantum coherence oscillations}  
An isomeric state $\psi_0$ of a quantum particle in an asymmetric DWP has a QCO resonance if it is  a linear superposition of two quasi-degenerate 
eigenstates  $\psi_d$, $\psi_u$ of the Hamiltonian, with energies $E_d \approx E_u$, such that $\Delta= \vert E_u - E_d \vert$ is  much smaller than the  average level spacing. In this case  the localization probability in the stable
well, $\rho(t)$, oscillates according to the law  
$\rho(t)= [1- \cos(\pi t/T_{max})]/2$, with $T_{max}= \hbar \pi
/ \Delta$. In most physical situations the resonance   
conditions are not fulfilled, but particularly interesting
are the systems where the potential depends on a parameter which may 
be changed continuously until there is resonance\footnote{Resonances in a three-well potential of variable width are presented in the application 2349433, {\it Long distance quantum transfer}, available at http://cipo.gc.ca}.  For  STM 
this parameter can be the surface-tip bias voltage U. \\ \indent 
The present estimates are based on    
the potential function given in ref. \cite{walk} for 
Xe atoms in a fixed surface-tip geometry. 
This potential  consists in a DWP term produced by the binding 
interaction  energy, $V_p(x)$, (Fig. 1, dashed line) and the  dipole term, 
$
V_d(x)= - U  \mu_0  \{ \lbrack  0.3+0.7 
(w+x)^4/L^4 \rbrack^{-1} 
- \lbrack 0.3+0.7 (w-x)^4/L^4 \rbrack^{-1} \} /2w $, where
$\mu_0=0.3$ D,
 (1D=3.335$ \times 10^{-30}$ C$\cdot$m), $w=2.2$ \AA,  and $L=1.56$ \AA.
The X-axis is chosen normal to the surface, with the origin 
at the barrier top of the term $V_p$. This term is taken as a fourth order polynomial, with the  metastable minimum $V_0=V_p(x_0)=-12.7$ meV   at $x_0=-0.7 $ \AA, near surface, the  barrier top $V_b=V_p(x_b)=0.45$ meV at $x_b=0$, and the 
stable  minimum $V_g = V_p(x_g) = -23.2$ meV at  $x_g= 0.89$ \AA, 
close to the tip.   At small bias  the total potential $V(x)= V_p(x) +V_d(x)$ remains a DWP, with the coordinates $x_0,x_b,x_g$ of the extrema close to the values given above. For negative bias of the surface the barrier height $E_b=V_b-V_0$ decreases, and at $U \approx -1.2 $ V the  isomeric minimum dissapears. \\ \indent 
The CM wave function for the metastable ground state of
a Xe atom localized initially near surface is well approximated
by the Gaussian wave packet  
$\psi_0(x)= (c_0 / \pi)^{1/4} e^{-c_0
(x-x_0)^2 /2} $,  where $c_0 = M \omega_0 / \hbar $,
$M$ is the Xe mass, and 
$\omega_0 = \sqrt{V'' (x_0)/M}$ is the classical oscillation
frequency at $x_0$  in the harmonic approximation. \\ \indent
If there are no decoherence factors, the wave function at the moment t is
\begin{equation}
\psi(x,t) = e^{-i \hat{H} t / \hbar} \psi_0(x)~~, ~~~~~~ \hat{H}=
- \frac{\hbar^2}{2 M} \frac{ \partial^2}{\partial x^2}
+V_p(x) +V_d (x)~~, 
\end{equation}   
and can be obtained by integrating the TDSE 
\begin{equation}
i \hbar \frac{ \partial \psi}{\partial t} = \hat{H} \psi
\end{equation}
with $\psi(x,0)= \psi_0(x)$ as initial condition. \\ \indent
The TDSE can be solved numerically in a spatial grid $\{ x_k \}$, k=1,N, 
using the leap-frog
method \cite{hym}. However, the computation becomes much faster 
if the discrete form of Eq. (2) is integrated as  
a Hamilton system of equations. Thus,
if $u_k (t) \equiv Re ( \psi(x_k,t) )$ and 
$v_k (t) \equiv Im ( \psi(x_k,t ))$ denote
the real, respectively the imaginary part of the 
wave function $\psi(x,t)$  at the grid point $x_k$, then Eq. (2) becomes
\begin{equation}
2 \hbar \dot{u}_k =  \frac{ \partial {\cal H} }{ \partial v_k} ~~~~~~~~~~
2 \hbar \dot{v}_k = - \frac{ \partial {\cal H} }{\partial u_k}~~,
\end{equation}
with 
\begin{equation}
{\cal H} = \sum_{k=1}^N u_k  (\hat{H} u)_k + v_k (\hat{H} v)_k~~,
\end{equation}
$$
(\hat{H}y)_k = - \frac{ \hbar^2}{2M \ell^2} \lbrack \frac{y_{k+3}+ y_{k-3}}{90}
-3 \frac{y_{k+2}+y_{k-2}}{20} + 3 \frac{y_{k+1}+y_{k-1}}{2} - \frac{49}{18}
y_k \rbrack +
$$
$$
+ \lbrack V_p(x_k) + V_d(x_k) \rbrack y_k~~. 
$$
In the present application the Hamiltonian system of Eq. (3)  was defined considering $N=321$ spatial grid points equally spaced by $\ell=0.01$ \AA~ within the interval $[ x_{min}, x_{max} ]$= [$-1.2$ \AA, 2 \AA ]. For a fast integration 
can be used the D02BAF routine of the NAG library \cite{nag}, with the time step $dt=6.58 \times 10^{-2} $ ps. The localization probability in the stable  well of $V(x)$ is $ \rho (t) = \int_{x_{b}}^{x_{max}}\psi^*(x,t) \psi(x,t) dx $, and  within the time interval [0,20 ps] attains a maximum which is  represented in Fig. 2(A) as a function of the bias voltage $U$. The peaks indicate the QCO resonances. The first appears at $U = -1.141 $ V for the potential  $V$ represented in Fig. 1 by solid line.  This resonance corresponds to $\rho(t)$ shown  in Fig. 2(B), with the 
maximum at  the moment  $T_{max}=14.37$ ps.  In Fig. 1 can be seen
the wave function at $t=0$, ($\psi_0$, dotted line), and at $t=T_{max}$, 
($\psi_M$, solid line). 
\section{The coherent tunneling}  
During QCO the wave-packet changes its shape, and therefore the tunneling
is not coherent. The issue of coherent tunneling may receive an answer
from the study of the evolution pattern for wave functions constrained to be
Gaussian all the time, as in the quantum molecular dynamics \cite{qmd}.    
The constrained dynamics will be obtained by a 
time-dependent variational calculation within the 
trial manifold of the Gaussian wave packets with variable centroid
and width. This manifold contains the isomeric ground state, and  
is represented by a combination between coherent and squeezed states, 
\begin{equation}
 \psi_c (z,s) =  e^{(z b^\dagger - z^* b)} e^{(s b^\dagger
b^\dagger- s^* b b)/2}  \psi_0~~, ~~~~~z= \alpha+i \beta,~ s = \rho e^{-2 i \phi}
~~,
\end{equation}
with  $(\alpha,\beta,\rho,\phi)$ real parameters, and
$b = \sqrt{ M \omega_0 / 2 \hbar} (x -x_0 + \hbar \partial_x/M \omega_0)$
the Dirac-Fock annihilation operator for $\psi_0$.
In terms of the variables 
$x_c = x_0+ \alpha \sqrt{ 2 \hbar / M \omega_0} $,
$p_c=  \beta \sqrt{ 2 \hbar M \omega_0} $, $v= (\sinh \rho)^2$,  
the equations of motion obtained from
the variational equation $\delta \int dt \langle \psi_c \vert
i \hbar \partial_t - \hat{H} \vert \psi_c \rangle=0$ are 
\begin{equation}
\dot{x_c}  = \frac{ \partial \langle \psi_c \vert \hat{H}
\vert \psi_c \rangle }{\partial p_c}
~~~~~~
\dot{p_c}  = - \frac{ \partial \langle \psi_c
\vert \hat{H} \vert \psi_c \rangle }{\partial x_c}~~,
\end{equation}
\begin{equation}
\hbar \dot{\phi}  = \frac{ \partial \langle \psi_c \vert \hat{H}
\vert \psi_c \rangle }{\partial v }
~~~~~~
\hbar \dot{v} = - \frac{ \partial \langle \psi_c
\vert \hat{H} \vert \psi_c \rangle }{\partial \phi }~~.
\end{equation}
Eqs. (6)  describe the motion of the CM coordinate and momentum
$x_c$, $p_c$ of the Xe atom, while Eqs. (7) provide the 
evolution of the localization width around $x_c$, 
$\sigma^2_x \equiv  \langle \hat{K} \rangle  = \langle x^2 \rangle - \langle x \rangle^2= [1/2 + v + \sqrt{v(v+1)} \cos 2 \phi]/c_0$. To express the   average $\langle \psi_c \vert \hat{H} \vert \psi_c \rangle$ in analytical form,  the potential $V(x)$  was replaced by a fourth order interpolation polynomial, 
$ V_{pol}(x)$, with coefficients determined by fit. \\ \indent 
At the first resonance ($U=-1.141$ V),  the best fit of the potential
$V(x)$ (Fig. 1, solid line) is given by a
quartic polynomial  $V_{pol}(x)$  (Fig.  3(A), solid line), with the 
extremum values of 0.8 meV, 1.85 meV and $-48.88$ meV at the points
$x_0=-0.47$ \AA,  $x_b=-0.19$ \AA, and $x_g=1.01$ \AA, respectively.
The classical oscillation frequency at $x_0$ which defines $\psi_0$ 
in Eq. (5)  is $\omega_0 = 1.95 $ ps$^{-1}$.  
Using these results,  $\hat{H}$ at resonance is well approximated by  $ \hat{H}_{pol} = - \hbar^2 \partial_x^2/2M + V_{pol} (x)$, and $V_{av} (x) \equiv \langle \hat{H}_{pol}\rangle  (x_c=x,p_c=0,v,\phi)$  has the role of effective potential for the CM dynamics of the Gaussian wave packet $\psi_c$.   This function is represented for $\phi=0$ and $v=0,0.1,0.5,1,1.5,2$ in Fig. 3(A) by dashed lines.
If $v=0$, then $V_{av}(x)$ is also a DWP, with the extremum
values of 1.57 meV, 1.97 meV and  $-46.8$ meV at $-0.425$ \AA,
$-0.215$ \AA, and 1 \AA, respectively. Thus, for a Gaussian
wave packet the effective potential is not the same as for the classical
particle, and the top of the barrier has a small shift to higher energy. \\ \indent
When $x_c=x_0$ and $p_c=v=\phi=0$, the trial function $\psi_c$ reduces to  
the isomeric ground state $\psi_0$ of $\hat{H}_{pol}$. The quasi-classical
dynamics of a Xe atom which is initially in this state was obtained by
integrating the system of Eqs. (6),(7) using the routine
D02BAF of the NAG library,  and the trajectory $x_c(t)$ is 
represented in Fig. 3(B) by dashed line. For comparison, the exact
wave function  $ \psi(t) =  \exp(-i t \hat{H}_{pol}/ \hbar ) \psi_0 $ was calculated
by solving the TDSE, and the corresponding expectation value $x_{QCO}(t) = \langle \psi (t) \vert x \vert \psi (t) \rangle$ is represented in Fig. 3(B) by solid line. Initially $x_c(t)$ and $x_{QCO}(t)$ are close, but in time $x_{QCO}(t)$ extends across the barrier, while $x_c(t)$ oscillates in the  isomeric well with small amplitude\footnote{Numerical estimates indicate that if the "squeezing" dissipation term provided by the coupling operator $\hat{K}= (x- \langle \psi \vert x \vert \psi \rangle )^2$ is included, and $\psi (t)$ is obtained by integrating the non-linear TDSE $i \hbar \partial_t \psi= (\hat{H}_{pol}+ \kappa \hat{K} \partial_t \langle \psi \vert \hat{K} \vert \psi \rangle) \psi $, then for $\kappa \sim 50 \hbar /$ \AA$^4$ any significant changes in the shape of $\psi$ are suppressed, and $\langle \psi (t) \vert x \vert \psi (t) \rangle$ remains close to $x_c(t)$. }. \\ \indent
The oscillation amplitude of $x_c$ increases with the energy 
$E_c = \langle \psi_c \vert \hat{H}_{pol} \vert \psi_c\rangle$, but the
trajectory remains confined in the isomeric well until
 $E_c$ reaches a narrow interval very close to the barrier top
of  $V_{av}$,  when a "switching" mode appears.
For an orbit with the initial conditions $x_c \sim -0.25$ \AA,
$p_c=v=\phi=0$, the oscillations in the isomeric well are interrupted by a
sudden jump to the stable well, where the particle keeps oscillating without return. This interesting behaviour resembles the experimental situation, because the observed atom switching occurs with a rate proportional to a power of the current, and could be explained by the electron heating effects \cite{ei, walk}. 
Here, this  mode is due to a small valley connecting the two wells which
appears  near the barrier  top of $V_{av}$  in the squeezing dimension.  
Thus, the escape to the stable well becomes possible
when the four dimensional orbit ${\cal C}_t  \equiv (x_c,p_c, v, \phi)(t)$ 
is directed along this valley.  The projection of the switching mode
orbit ${\cal C}$ on the CM phase space plane 
is pictured in Fig. 4(A), while the corresponding trajectory plot,  $x_c(t)$,
is shown in Fig. 4(B).   
\section{Conclusions}  
The TDSE calculations indicate that if quantum coherence is preserved, then elastic tunneling of the Xe atom can appear only at certain resonant
values of the bias voltage. At resonance the atom oscillates 
between a localized state on the surface ($\psi_0$), and a state 
close to tip ($\psi_M$), with a frequency attaining the maximum 
($\sim 1/2 T_{max}=34.8$  GHz)  at the first resonance  ($U \sim -1.141$ V). 
Thus, the exact treatment of the atom tunneling in the STM potential predicts an oscillatory behavior which cannot explain the irreversible switching mode. The observation of irreversible transfer with a constant rate  
indicates that in the present experiments the
phase coherence is destroyed by the voltage pulse, or by other
external factors, like dissipation.  According to the quasi-classical results, 
the switching might be explained by a decoherence mechanism  producing
excitation and partial localization. 
\\ \indent
The atomic QCO in STM might be observed as oscillations of the junction
impedance, which changes when the atom moves from surface to the tip \cite{ei}. Therefore, accurate measurements of the resonant bias voltages and of the transfer time-scale could provide significant insight on the mechanism of decoherence and localization at the atomic level.
 
{\bf Figure Captions} \\[.5cm]
Fig. 1. The potential $V$ at $U=-1.141$ V (solid), 
$U=0$ (dash),  and the wave functions $\psi_0$, $\psi_M$   
 (in units of \AA$^{-1/2}$)  as a function of distance. \\ 
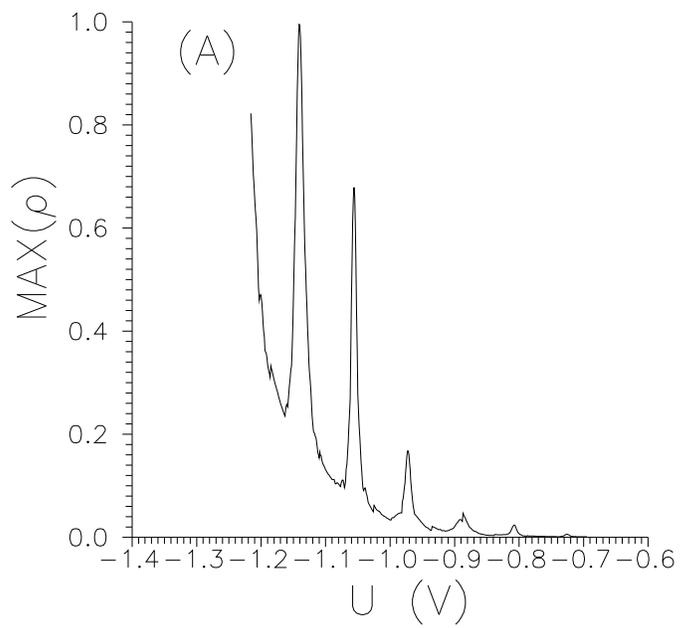
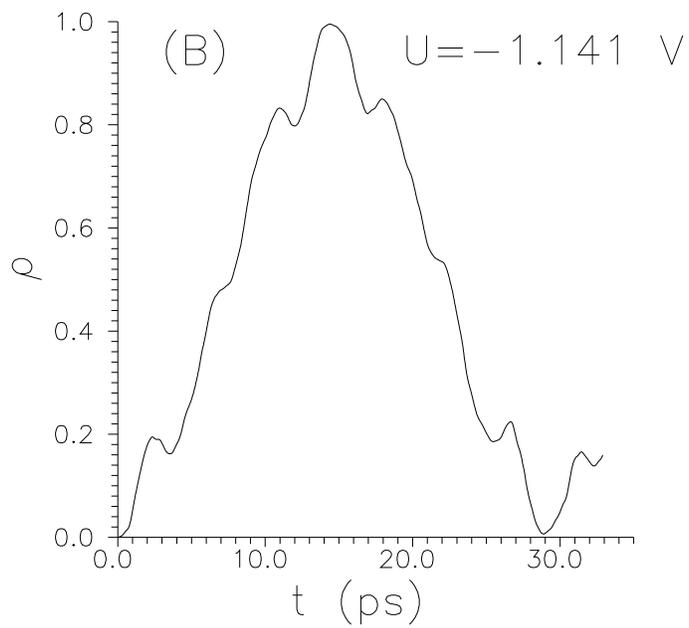
Fig. 2. The maximum value attained by $\rho$  within
20 ps as a function of the bias voltage (A) and
$\rho$  as a function of time at the first resonance (B).  \\
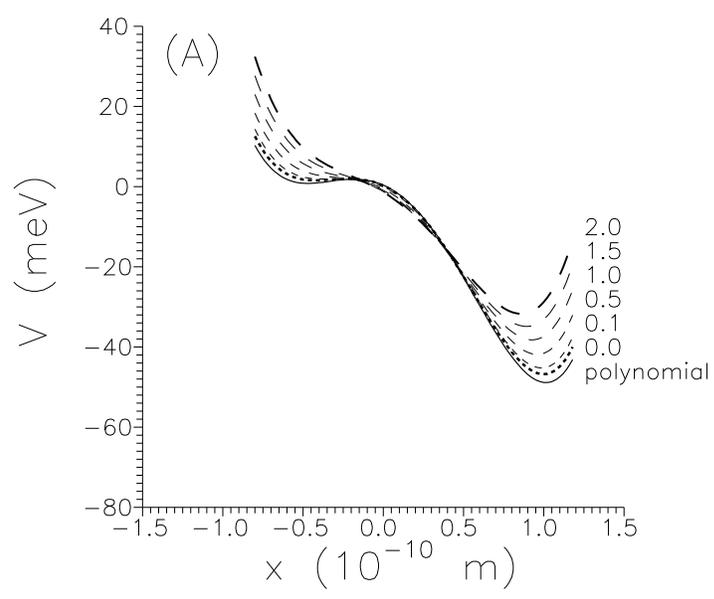
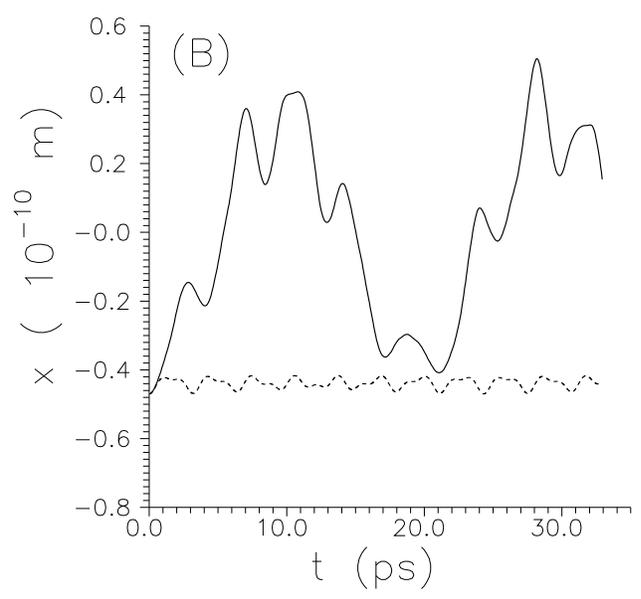
Fig. 3. The polynomial potential $V_{pol}$ (solid) and the average potential 
$V_{av}$ (dash) for $v=0,0.1,0.5,1,1.5,2$, as a function of distance (A);
$x_{QCO}$ (solid) and $x_c$ (dash) as a function of time (B). \\
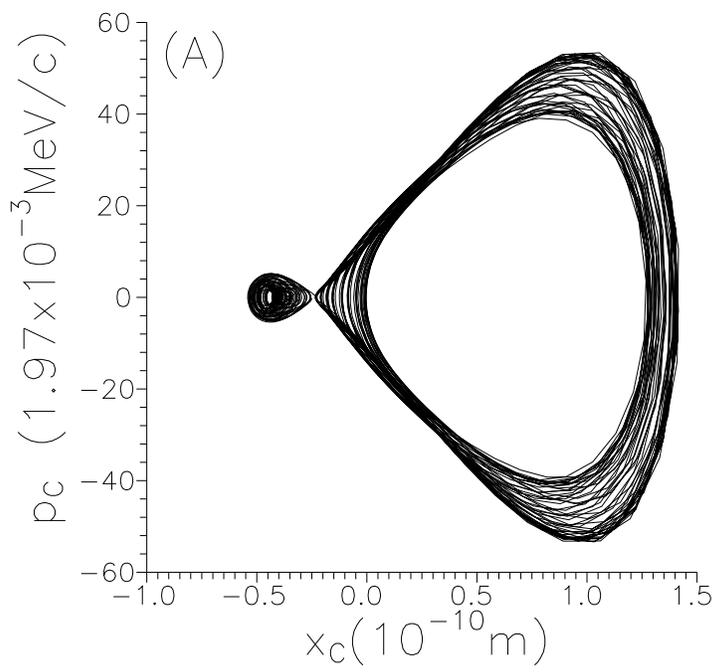
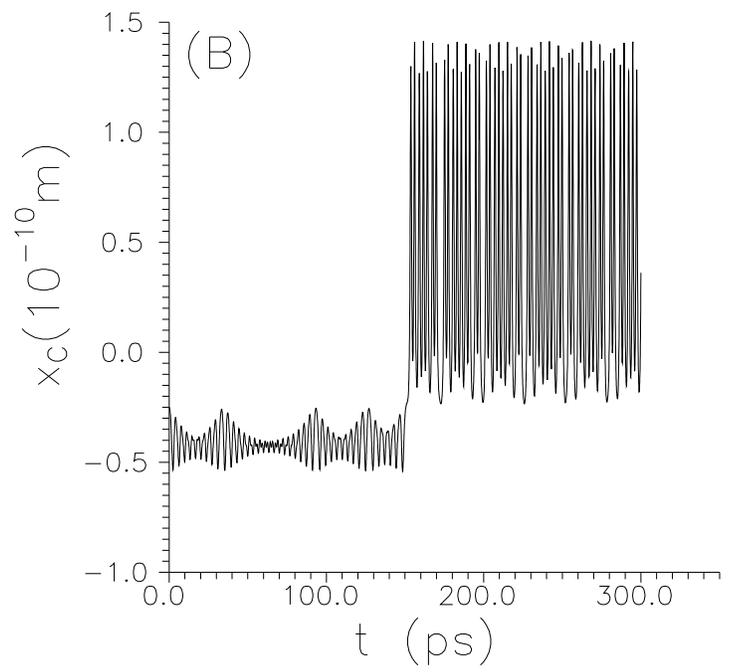
Fig. 4.  The switching mode: phase-space orbit of the CM motion (A) and
the CM coordinate as a function of time (B). 
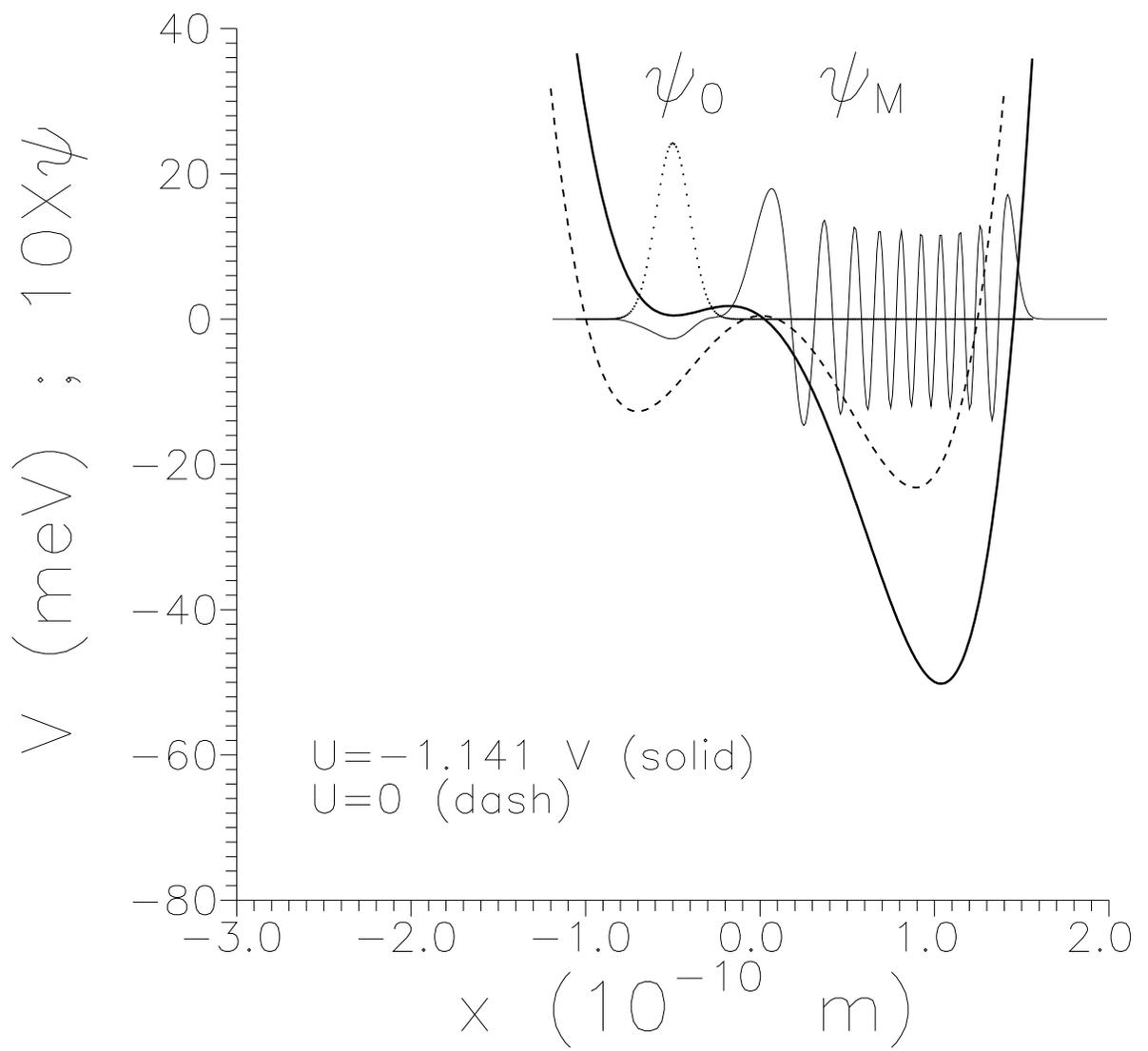
\end{document}